# Revealing the terahertz-laser velocity effect during air filamentation via travelling-wave-antenna model

Jiajun Yang,[1] Xiaofeng Li,[1] Linlin Yuan,[1] Li Lao,[2] and Jiayu Zhao[1,3,a]

**AFFILIATIONS**

[1]Terahertz Technology Innovation Research Institute, Terahertz Spectrum and Imaging Technology Cooperative Innovation Center, Shanghai Key Lab of Modern Optical System, University of Shanghai for Science and Technology, Shanghai 200093, China

[2]Tera Aurora Electro-optics Technology Co., Ltd, Shanghai 200093, China.

[3]Shanghai Institute of Intelligent Science and Technology, Tongji University, Shanghai 200092, China

[a] *zhaojiayu@usst.edu.cn*

**ABSTRACT**

During femtosecond laser filamentation in air, the velocity ratio ($K$) between the terahertz (THz) phase velocity and the laser group velocity plays a crucial role in THz waves generation. However, $K$ is typically assumed to be unity and its impact has been long overlooked due to the more attention paid to the more easily controlled filament length. Here, we investigate the obscured contribution of $K$ to the THz radiation characteristics by using the improved travelling-wave-antenna (TWA) model. It has been found that, under both single- and two-color laser pumping schemes, $K$ significantly determines the far-field spatial distribution of forward or backward THz radiation, as well as a transition from Bessel- to Cherenkov-type THz emission patterns. These results establish the TWA model as a reliable theoretical tool for studying the mechanisms of THz beam shaping via the designed $K$. Moreover, for cases of $K$ not being controlled, its value can also be inferred by the proposed TWA model, which could be an effective method to confirm whether the laser ionization front is superluminal or subluminal compared with the generated THz waves.

**Introduction.** Over the past two decades, terahertz (THz) technology has rapidly advanced as a cutting-edge research field, achieving remarkable progress in both fundamental studies and practical applications. Meanwhile, a variety of THz sources have emerged, including quantum cascade lasers [1], photoconductive antennas [2], laser-pumping nonlinear crystals [3], laser ionization of solids [4] and transparent media (e.g., air) [5–8]. Among these THz sources, the air plasma during femtosecond laser filamentation has attracted significant attention in recent years due to its distinctive advantages, such as broad bandwidth and high output intensity [9,10]. And in this aspect, the ratio between the THz phase velocity and the laser group velocity (denoted as $K$) is an essential parameter, no matter in cases of the transition-Cherenkov THz radiation [11–16] and the flying-focus technique [17, 18] under single-color fields, or in typical THz radiation profiles such as Bessel [19] and conical types [20, 21] under dual-color fields.

However, due to the tediousness and complexity in directly measuring the laser wavefront propagation velocity [17, 18, 22, 23], it remains challenging to easily determine the value of $K$ in experiments. In this context, previous studies tended to simplify their theories by setting $K$ as 1 [11–15, 20, 21]. Meanwhile, the plasma filament length ($l$) closely relates to the THz radiation characteristics and exhibits a similar variation trend with $K$: under tighter focusing conditions, as $l$ decreases, $K$ decreases accordingly, causing the THz radiation to deviate from the laser axis; conversely, as $l$ increases, $K$ increases, directing the THz radiation along the axis. As a result, current studies have primarily focused on manipulating $l$ [5, 24, 25], while the role of $K$ has received relatively limited attention. It is also worth noting that due to the uncertainty of the filament's geometric length, imprecision of the $l$ value may obscure the contribution of $K$ in the control of THz beam profiles.

Very recently, with the advancement of laser wavefront control techniques, the significance of $K$ has attracted increasing attention. For instance, under single-color laser field conditions, Fu, et al., [17, 18] demonstrated the controllability of $K$ by using the flying focus method to modulate the laser wavefront propagation velocity, and further verified the $K$'s critical role in determining the energy distribution of THz radiation in the far field. Specifically, a positive $K$ directs the THz main lobe along the laser path, while a negative $K$ reverses it. Moreover, as the magnitude of $K$ increases, the THz radiation direction gradually aligns more closely with the laser axis. Furthermore, under dual-color conditions, Zhang, et al., [19] combined a planoconvex lens and an axicon to generate the plasma filament with a nonuniform axial $K$ distribution, which enabled the observation of far-field Bessel-type THz radiation. When the variation range of $K$ narrowed and gradually approached the constant, the THz beam profile transitioned from a Bessel- to a Cherenkov-like radiation pattern. Both the above findings suggest that $K$ is no longer a neglectable parameter.

Inspired by these reports, the present work systematically investigates the influence of $K$ on the THz radiation envelopes under the same framework of travelling-wave-antenna (TWA) concept [20, 21]. Several representative experimental configurations have been chosen, namely, a single filament in the one- [17, 18] or two-color field [19], double counter-propagating filaments [26], and laser-illuminated metal wires [27], respectively. Our results indicate that the TWA model can successfully reproduce the THz radiation patterns related to $K$ as reported in the existing literature, which hence has been demonstrated to be an effective theoretical tool on the $K$-control mechanisms of THz radiation.

**The TWA concept**. To establish the theoretical framework for this study, we adopted the unified TWA equation previously validated by our group [20, 21]:

$$\left|E(\omega,\theta,l,\phi_0)\right| \propto \left|\sin\theta\right| \cdot \left|\left[e^{i\phi_0} \frac{\sin[(\frac{kl}{2})(K-\cos\theta-\frac{\lambda}{2l_d})]}{(K-\cos\theta-\frac{\lambda}{2l_d})} - e^{-i\phi_0} \frac{\sin[(\frac{kl}{2})(K-\cos\theta+\frac{\lambda}{2l_d})]}{(K-\cos\theta+\frac{\lambda}{2l_d})}\right]\right| \quad (1)$$

which could consistently describe the THz radiation characteristics generated by both single- and dual-color laser fields. Here, $E(\omega,\theta,l,\phi_0)$ is defined as the THz electric field, $\omega$ is the THz angular frequency, $\theta$ is the angle relative to the laser axis, $l$ represents the plasma filament length, and $\phi_0$ is the initial phase difference between the fundamental laser and its second harmonic. Additionally, $k = \omega/c$ is the THz wave number in free space, and $K$ indicates the ratio between the THz phase velocity and the laser group velocity. The dephasing length $l_d$ is the distance over which the phase difference between the two colors shifts by $\pi$.

To facilitate our subsequent exploration of the THz field generated by the interference of counter-propagating filaments in the single-color laser field, we also derive a modified expression based on Eq. (1). Briefly, as the dephasing length $l_d$ approaches infinity, Eq. (1) simplifies to the single-color case [21]. Then, we adjust $\theta$ for the two counter-propagating filaments: for one filament, $\theta$ is replaced with $\theta + \pi/2$; and for the other, $\theta$ is replaced with $\theta - \pi/2$. The overall THz radiation is formed by $|E'(\omega,\theta,l)| = |E(\omega,(\theta + \pi/2),l) - E(\omega,(\theta - \pi/2),l)|$ as

$$|E'(\omega,\theta,l)| \propto |\cos\theta| \cdot \left| \frac{\sin\left(\frac{kl}{2}[K+\sin\theta)]\right)}{K+\sin\theta} + \frac{\sin\left(\frac{kl}{2}[K-\sin\theta]\right)}{K-\sin\theta} \right| \quad (2)$$

In following sections, Eqs. (1-2) will be used separately to study the $K$ effect on the THz beam profiles.

**Controlling the THz spatial distribution by modulating $K$ in the single-color case.** Recently, Fu, et al., [17, 18] precisely controlled the laser wavefront velocity in the field of THz radiation from a single-color laser filament, enabling the tunability of $K$ and the flexible shaping of the THz beam envelope. In order to compare with the reported results, we similarly started with a low frequency of 0.1 THz from a 1-mm-long single-color filament under different $K$ values (1:1.8, 1:1, 1:0.6, 1:-1.2 and 1:-0.8), and based on Eq. (1) with $l_d \sim \infty$, we calculated the THz radiation distribution as shown in Fig. 1(a). It can be seen that a positive $K$ directs the THz main lobe along the laser path (0°), while a negative $K$ reverses it. Moreover, as $|K|$ increases, the THz radiation direction gradually approaches the filament axis. These results are consistent with Refs. [17, 18]. We also increased the filament length (e.g., to 3 mm), and the TWA-calculated results (not shown) demonstrate the same THz radiation trend with that in Fig. 1(a) of this work and Fig. 3(b) of Ref. [17].

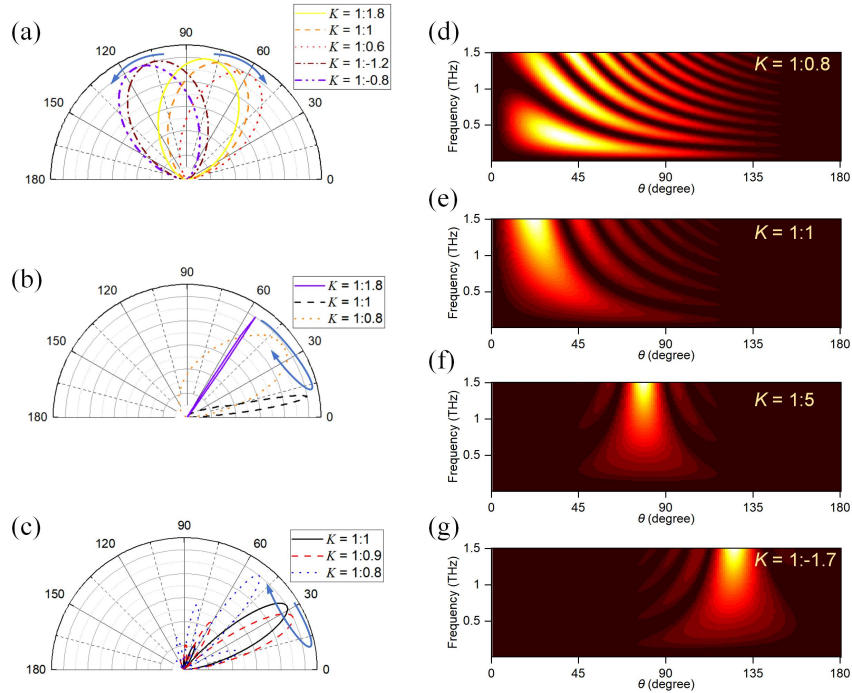

Fig. 1. THz radiation characteristics of a 1-mm-long single-color filament under different $K$ values. (a) At 0.1 THz. (b) For a broadband range (0.1-10 THz). (c) At 0.7 THz. (d-g) The THz distribution as functions of frequency and radiation angle.

Furthermore, based on the literature [17], we changed the above 0.1 THz to a broadband THz radiation (0.1-10 THz) with *K* values of 1:1.8, 1:1, and 1:0.8. It can be observed in Fig. 1(b) that, for *K* = 1:0.8, the THz profile exhibits a broad angular distribution, while for *K* = 1:1.8 or 1:1, the THz radiation is concentrated within a narrow angle, which facilitates a highly directional THz emission and detection. This behavior is also consistent with the results reported in Ref. [17]. However, it has not been clearly explained in Ref. [17] why as |*K*| increases, the THz direction initially moves closer to the laser axis and then deviates along the arrow in Fig. 1(b), displaying a different trend in the THz radiation angle compared to Fig. 1(a). About this point, we additionally analyzed the THz radiation at different frequencies beyond 0.1 THz. Taking 0.7 THz as an example, as shown in Fig. 1(c), one can see that as *K* increases, the main THz lobe gradually moves closer to the filament axis but weakens. Meanwhile, the side lobe increases, and then becomes dominant and form a new direction of the primary THz radiation along the arrow in Fig. 1(c). These phenomena could explain why under broadband conditions as in Fig. 1(b), the THz radiation angle for *K* = 1:0.8 is larger than that for *K* = 1:1.

More calculations at different THz frequencies (0-1.5 THz) are shown in Figs. 1(d-g), following Ref. [18]. Here, the THz spectral distributions are as a function of the radiation angle $\theta$ with different *K* values (1:0.8, 1:1, 1:5 and 1:-1.7). The first three figures (Figs. 1(d-f)) show that as *K* decreases from 1:0.8 to 1:5, the THz radiation becomes more concentrated in a narrower angular range, consistent with the trend in Fig. 1(b). And in Fig. 1(g) with *K* = 1:-1.7, the THz emission direction is opposite to the laser propagation, in agreement with Fig. 1(a) for negative *K* values. All these calculations in Fig. 1 based on the TWA model confirm the significant *K* effect on the profile control of THz beams. It is also worth noting that, since the TWA model doesn't consider the high-frequency decay of the THz radiation efficiency from a laser filament, the THz distributions towards 1.5 THz in Figs. 1(d-g) are stronger than those in Figs. 5(b-e) of Ref. [18]. To address this issue, we introduced a simple window function to Figs. 1(d-g), which then could well match the reported results as can be seen in Supplemental Material S1.

**Controlling the Bessel-type THz spatial distribution by modulating *K* in the two-color case.** To further validate the *K*'s role, we examined the effect of a non-uniform *K* along the dual-color plasma filament on the THz radiation, following Ref. [19]. Specifically, based on Eq. (1), THz radiations were calculated at 0.4 THz from a 55-mm-long filament with a 24.6-mm dephasing length ($l_d$), where the laser ionization front velocity increased from 1.006*c* to 1.016*c* (*c* is the light speed in vacuum) along the plasma filament given by a combined focusing of lens and axicon. Thus, *K* varied between 1:1.006 and 1:1.016. And we also changed this *K* range into a large one as 1:1.004-1:1.019, or a smaller one as 1:1-1:1 (constant) along the filament.

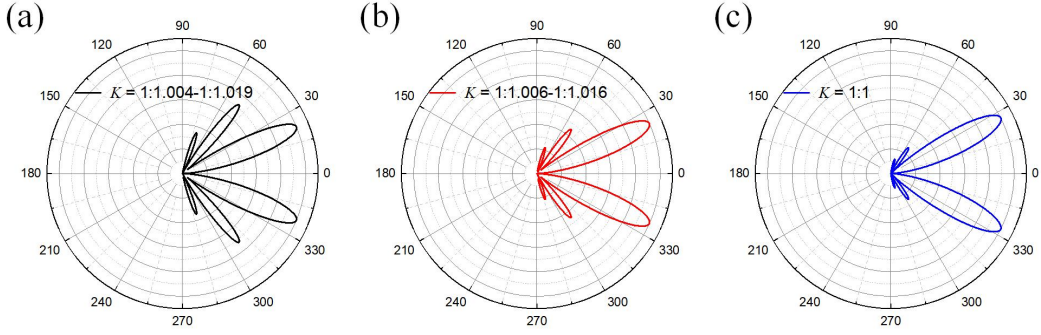

Fig. 2. THz radiation patterns in the two-color case with a *K* variation along the plasma filament: (a) from 1:1.004 to 1:1.019, (b) from 1:1.006 to 1:1.016, (c) fixed at 1.

Corresponding calculation results are shown in Fig. 2, inside which one can see the multiple side lobes characteristic of a Bessel-type THz beam in Fig. 2(a), resulted from the spatial interference from plasma sources with different *K* values along the laser filament. Then, as the *K* range was reduced as shown in Fig. 2(b), the interference effect weakened, reducing the side lobes while still maintaining some Bessel beam features. Finally, the interference effect is no longer considered when *K* becomes constant as 1 along the filament. In this case, as shown in Fig. 2(c), the THz radiation pattern transitions from a multiple side lobes structure to a conical propagation within a specific angle, displaying the well-known transition-Cherenkov-like radiation. The above TWA-predicted results confirm that the axial distribution of *K* along the filament plays a key role in steering the THz beam, consistent with the literature [19].

**Determination of the *K* value via the TWA model.** In previous sections, the TWA model has been validated to accurately predict the THz radiation characteristics with designed *K* values under both single- and dual-color field conditions. In this section, for cases of *K* not being controlled, its value has been inferred by fitting the experimental/simulated THz radiation results with the proposed TWA model, which could be an effective method to determine *K* and the laser ionization front velocity in intense field experiments. However, before computing the *K* value, other parameters in the TWA formula should be firstly decided, among which the filament length *l* might be the fuzziest one since the plasma filament is not a geometrical cylinder. On the other hand, intrinsic correlations come between *l* and *K* as mentioned in the introduction. Therefore, imprecision in the *l* value may obscure the contribution of *K*, which then cannot be accurately determined. Now, the question becomes in which case *l* can be easily decided, and two exampled cases are as follows.

**(1) THz radiation generated by two counter-propagating filaments.** The first example involves a counter-propagating filament array, where two laser pulses travel in opposite directions and interact. In this case, the effective *l* for THz generation is evaluated by the laser interaction length, which corresponds to the transmission distance of the laser pulses within the duration. For example, in Ref. [26], when two 200-fs laser pulses collide in a counter-propagating manner as shown by the blue arrows in the inset of Fig. 3, *l* can be estimated as (200-400) fs × 3 × $10^8$ m/s, resulting in *l* = 60-120 μm. Then, we can employ the TWA model (Eq. (2)) to calculate the *K* value by fitting the 2.1-THz experimental data in Ref. [26], which is primarily emitted perpendicular to the laser propagation axis as shown as red circles in the inset of Fig. 3.

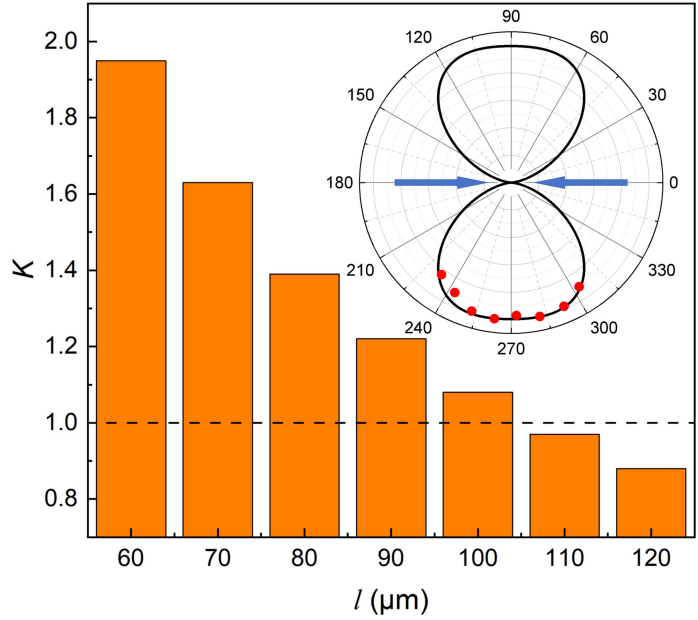

Fig. 3. In case of a counter-propagating filament array, $K$ as a function of $l$ when fitting the THz data in the literature [26] (inset) with the TWA model.

When the TWA-fitting line (black) closely follows the THz data (red) in the inset of Fig. 3, the $K$ values are obtained and plotted in Fig. 3. It can be seen that $K$ decreases as $l$ increases, but its variation remains mostly above 1 (horizontal dashed line in Fig. 3). This indicates that in Ref. [26], the laser propagation didn't exceed the THz speed. Normally, this superluminal/subluminal issue is fuzzy or difficult to be decided in a easy way. Now, the TWA model has offered a new solution. Besides, more significance of this result is provided in Discussion.

**(2) THz radiation generated by laser-illuminated metal wires.** The second example involves THz radiation from a laser-illuminated metal wire [27], and in this case, the wire length (6, 30 or 800 μm) is considered as the effective $l$ for the off-axis 12.6-THz radiation. To reproduce this result, we also employed the TWA model (Eq. (1) for single-color case) because of the inherent similarity between plasma-filament-based and metal-wire-based THz sources. With identical parameters from Ref. [27] and by adjusting the $K$ value, we achieved the best reproductions as shown in the inset of Fig. 4. One can see that for shorter metal wires (6 and 30 μm), the THz radiation is primarily distributed perpendicular to the laser propagation direction. As $l$ increases to 800 μm, its direction tilts toward the laser axis, exhibiting stronger directionality. These trends are generally consistent with the results reported in Ref. [27] (see also Supplemental Material S2).

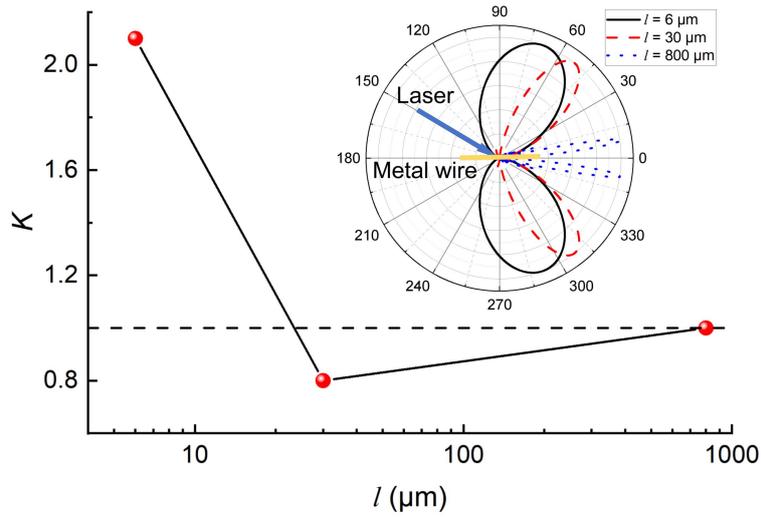

Fig. 4. In case of a laser-illuminated metal wire, $K$ as a function of $l$ when reproducing the THz data in the literature [27] (inset) with the TWA model.

The variation of $K$ with $l$ is shown in Fig. 4, where the $K$ values are 2.1, 0.8 and ~1 for $l$ = 6, 30 and 800 μm, respectively. These observations can be interpreted as that, the extreme short metal-wire length (6 μm) introduces large uncertainty into the TWA model calculation of the $K$ value. And this causes $K$, which should be less than 1 to promote the off-axis THz emission (black line in the inset of Fig. 4), to exceed 1 (2.1). Nevertheless, when the metal wire is sufficiently long (30 and 800 μm), the TWA model could then calculate reasonable $K$ values as below 1 (0.8) or approximately equal to 1, respectively, in much better agreements with the THz envelope features as off-axis and nearly on-axis radiations (red and blue lines in the inset of Fig. 4).

**Discussion.** Note that, the transition-Cherenkov radiation model [28] has been widely accepted to explain the far-field conical THz radiation generated by single-color femtosecond laser filaments in gases. And the theoretical premise of this model requires that the speed of the charged particle exceeds the THz wave in the same medium. However, the same research team later pointed out in Ref. [29] about their theory that "superluminality is not strictly necessary." This has raised a question: whether the Cherenkov-like conical THz radiation observed in common-lens-focusing experiments originates from a superluminal or subluminal laser ionization front? To answer this question, remind ourselves that in Fig. 3, we performed THz calculations under the conventional lens focusing condition with a focal length of 2 inches for each filament from the counter-propagating filaments [26]. And the $K$ values were achieved as basically greater than 1, indicating that the propagation speed of the laser ionization front is slower than THz waves in air.

This result is meaningful, since it not only is a simple method to judge a superluminal or subluminal laser ionization front benefited from the proposed TWA model, but also could clarify the aforementioned fuzzy experimental picture that, the laser front is indeed subluminal and it can still effectively produce a far-field conical THz radiation from a single filament (and 90-degree off-axis THz radiation from two counter-propagating filaments). Hence, the TWA model is proved to be versatile to both manage (Fig. 1-2) and predict (Fig. 3-4) $K$ in laser-induced THz generation experiments. On the other hand, the TWA model has also been successfully applied to study the THz radiation phenomena in non-gaseous transparent media (e.g., atomic cluster and water film) [20], further highlighting the accuracy and applicability of this model in revealing the physical mechanisms of THz emission in related fields.

**Conclusion.** Based on the re-worked TWA model, we systematically studied the influence of $K$ on THz radiation characteristics. First, under single-color field, the modulation of THz emission directionality by varying $K$ was revealed. Subsequently, under dual-color field, the effect of nonuniform $K$ distributions along the filament on far-field THz radiation is explored, unveiling the transition from a Bessel- to a Cherenkov-type THz radiation. Furthermore, by analyzing cases involving counter-propagating filaments and laser-illuminated metal wires, the variation trend of $K$ with effective filament length was deduced, confirming the experimental fact that the laser ionization front propagates at subluminal velocities under 2-inch focal length conditions. Through these investigations, it is demonstrated that the suggested TWA model could accurately capture the significant $K$ effect on THz radiation characteristics in situation of various experimental configurations, verifying its effectiveness as a theoretical tool for studying the underlying mechanisms associated with $K$, and laying a foundation for future precise control of THz beam profiles based on $K$ engineering.

## SUPPLEMENTARY MATERIAL
See the supplementary material for additional information

## ACKNOWLEDGMENT
This work was supported in part by National Natural Science Foundation of China (61988102, 62435010, 62335012), National key research and development program (2023YFF0719200, 2022YFA1404004), Shanghai Rising-Star Program (22QC1400300), 111 project (D18014).

## AUTHOR DECLARATIONS
**Conflict of Interest**

The authors declare no conflicts of interest.

**Author Contributions**

**Jiajun Yang** & **Xiaofeng Li**: Writing – original draft (equal); Writing – review & editing (equal); Software (equal). **Linlin Yuan**: Investigation (equal); Writing – original draft (equal); Writing – review & editing (supporting); Software (equal). **Li Lao**: Writing – review & editing (equal); Resources (supporting). **Jiayu Zhao**: Conceptualization (lead); Methodology (equal); Writing – original draft (lead); Writing – review & editing (lead).

## DATA AVAILABILITY
Data underlying the results presented in this paper are not publicly available at this time but may be obtained from the authors upon reasonable request.

## REFERENCES

[1]A. Khalatpour, A. Paulsen, C. Deimert, et al., "High-power portable terahertz laser systems," Nat. Photonics **15**, 16 (2021).

[2]X. Ropagnol, Z. Kovács, B. Gilicze, et al., "Intense sub-terahertz radiation from wide-bandgap semiconductor based large-aperture photoconductive antennas pumped by UV lasers," New J. Phys. **21**, 113042 (2019).

[3]B. Zhang, Z. Ma, J. Ma, et al., "1.4-mJ high energy terahertz radiation from lithium niobates," Laser Photonics Rev. **15**, 2000295 (2021).

[4]G. Liao, Y. Li, H. Liu, et al., "Multimillijoule coherent terahertz bursts from picosecond laser-irradiated metal foils," Proc. Natl. Acad. Sci. U. S. A. **116**, 3994 (2019).

[5]Z. Zhang, Y. Chen, M. Chen, et al., "Controllable terahertz radiation from a linear-dipole array formed by a two-color laser filament in air," Phys. Rev. Lett. **117**, 243901 (2016).

[6]L. Zhang, W. Wang, T. Wu, et al., "Strong terahertz radiation from a liquid-water line," Phys. Rev. Appl. **12**, 014005 (2019).

[7]A. Koulouklidis, C. Gollner, V. Shumakova, et al., "Observation of extremely efficient terahertz generation from mid-infrared two-color laser filaments," Nat. Commun. **11**, 292 (2020).

[8]A. Balakin, J. Coutaz, V. Makarov, et al., "Terahertz wave generation from liquid nitrogen," Photonics Res. **7**, 678 (2019).

[9]H. Roskos, M. Thomson, M. Kreß, et al., "Broadband THz emission from gas plasmas induced by femtosecond optical pulses: From fundamentals to applications," Laser Photonics Rev. **1**, 349 (2007).

[10]M. Clerici, M. Peccianti, B. Schmidt, et al., "Wavelength scaling of terahertz generation by gas ionization," Phys. Rev. Lett. **110**, 253901 (2013).

[11]Y. Liu, A. Houard, B. Prade, et al., "Amplification of transition-Cherenkov terahertz radiation of femtosecond filament in air," Appl. Phys. Lett. **93**, 051108 (2008).

[12]A. Houard, Y. Liu, B. Prade, et al., "Strong enhancement of terahertz radiation from laser filaments in air by a static electric field," Phys. Rev. Lett. **100**, 255006 (2008).

[13]S. Mitryukovskiy, Y. Liu, B. Prade, et al., "Coherent interaction between the terahertz radiation emitted by filaments in air," Laser Phys. **24**, 094009 (2014).

[14]S. Mitryukovskiy, Y. Liu, B. Prade, et al., "Coherent synthesis of terahertz radiation from femtosecond laser filaments in air," Appl. Phys. Lett. **102**, 221107 (2013).

[15]S. Mitryukovskiy, Y. Liu, B. Prade, et al., "Effect of an external electric field on the coherent terahertz emission from multiple filaments in air," Appl. Phys. B **117**, 265 (2014).

[16]G. Hu, B. Shen, A. Lei, et al., "Transition-Cherenkov radiation of terahertz generated by super-luminous ionization front in femtosecond laser filament," Laser Part. Beams **28**, 399 (2010).

[17]S. Fu, B. Groussin, Y. Liu, et al., "Steering laser-produced THz radiation in air with superluminal ionization fronts," arXiv preprint **2407**, 18579 (2024).

[18]S. Fu, B. Groussin, Y. Liu, et al., "Steering laser-produced THz radiation in air with superluminal ionization fronts," Phys. Rev. Lett. **134**, 045001 (2025).

[19]Z. Zhang, J. Zhang, Y. Chen, et al., "Bessel terahertz pulses from superluminal laser plasma filaments," Ultrafast Science **2022**, 9870325 (2022).

[20]J. Zhao, Q. Wang, Y. Hui, et al., "Traveling-wave antenna model for terahertz radiation from laser-plasma interactions," SciPost Phys. Core **5**, 046 (2022).

[21]F. Zhu, J. Zhao, L. Lao, et al., "Unified framework for terahertz radiation from a single- or two-color plasma filament," Opt. Lett. **49**, 41 (2024).

[22]B. Guo, J. Sun, Y. Lu, et al., "Ultrafast dynamics observation during femtosecond laser-material interaction," International Journal of Extreme Manufacturing **1**, 032004 (2019).

[23]J. Sun, H. Cai, S. Zhang, et al., "Diagnostics of ionization front and plasma within femtosecond laser interferometer," Proc. of SPIE **10255**, 473 (2017).

[24]A. Gorodetsky, A. Koulouklidis, M. Massaouti, et al., "Physics of the conical broadband terahertz emission from two-color laser-induced plasma filaments," Phys. Rev. A **89**, 033838 (2014).

[25]Y. You, T. Oh, K. Kim, "Off-Axis phase-matched terahertz emission from two-color laser-induced plasma filaments," Phys. Rev. Lett. **109**, 183902 (2012).

[26]Y. Chen, Y. He, L. Liu, et al., "Interaction of colliding laser pulses with gas plasma for broadband coherent terahertz wave generation," Photonics Res. **11**, 1562 (2023).

[27]H. Zhuo, S. Zhang, X. Li, et al., "Terahertz generation from laser-driven ultrafast current propagation along a wire target," Phys. Rev. E **95**, 013201 (2017).

[28]C. D'Amico, A. Houard, M. Franco, et al., "Conical forward THz emission from femtosecond-laser-beam filamentation in air," Phys. Rev. Lett. **98**, 235002 (2007).

[29]C. D'Amico, A. Houard, S. Akturk, et al., "Forward THz radiation emission by femtosecond filamentation in gases: theory and experiment," New J. Phys. **10**, 013015 (2008).